\documentclass[reprint,prl,twocolumn,showpacs,superscriptaddress,floatfix]{revtex4-1}

\usepackage{hyperref}
\usepackage{latexsym}
\usepackage{graphicx}
\usepackage{amsmath}
\usepackage{amssymb}
\usepackage{color}
\usepackage{nicefrac}

\newcommand{\fig}{Fig.\ \ref}

\begin{document}

\title{Spin- and energy-dependent tunneling through a single molecule with intramolecular 
spatial resolution}
\author{Jens Brede}
\affiliation{Institute of Applied Physics, University of Hamburg, Germany}

\author{Nicolae Atodiresei}
\email{n.atodiresei@fz-juelich.de}
\affiliation{Institut f{\"u}r Festk{\"o}rperforschung and Institute for Advanced Simulation, Forschungszentrum J{\"u}lich, 52425 J{\"u}lich, Germany}
\affiliation{The Institute of Scientific and Industrial Research, Osaka University, Japan}

\author{Stefan Kuck}
\affiliation{Institute of Applied Physics, University of Hamburg, Germany}

\author{Predrag Lazi\'{c}}
\affiliation{Institut f{\"u}r Festk{\"o}rperforschung and Institute for Advanced Simulation, Forschungszentrum J{\"u}lich, 52425 J{\"u}lich, Germany}
\affiliation{Rudjer Boskovic Institute, Zagreb, Croatia}

\author{Vasile Caciuc}
\affiliation{Institut f{\"u}r Festk{\"o}rperforschung and Institute for Advanced Simulation, Forschungszentrum J{\"u}lich, 52425 J{\"u}lich, Germany}

\author{Yoshitada Morikawa}
\affiliation{The Institute of Scientific and Industrial Research, Osaka University, Japan}

\author{Germar Hoffmann}
\email{ghoffman@physnet.uni-hamburg.de}
\affiliation{Institute of Applied Physics, University of Hamburg, Germany}

\author{Stefan Bl{\"u}gel}
\affiliation{Institut f{\"u}r Festk{\"o}rperforschung and Institute for Advanced Simulation, Forschungszentrum J{\"u}lich, 52425 J{\"u}lich, Germany}

\author{Roland Wiesendanger}
\affiliation{Institute of Applied Physics, University of Hamburg, Germany}

\begin{abstract}

We investigate the spin- and energy dependent tunneling through a single organic molecule (CoPc) adsorbed on a ferromagnetic Fe thin film, spatially resolved by low-temperature spin-polarized scanning tunneling microscopy. Interestingly, the metal ion as well as the organic ligand show a significant spin-dependence of tunneling current flow. State-of-the-art \textsl{ab initio} calculations including also van-der-Waals interactions reveal a strong hybridization of molecular orbitals and surface $3d$ states. The molecule is anionic due to a transfer of one electron, resulting in a non-magnetic (S$=0$) state. Nevertheless, tunneling through the molecule exhibits a pronounced spin-dependence due to spin-split molecule-surface hybrid states.

\end{abstract}

\pacs{68.37.Ef, 31.15.E-, 75.70.-i, 75.50.Xx}
%73.20.-r 	Electron states at surfaces and interfaces
%75.70.-i 	Magnetic properties of thin films, surfaces, and interfaces
%31.15.E- 	Density-functional theory 
%68.37.Ef		Scanning tunneling microscopy (STM) in study of surface structure
%75.50.Xx 	Molecular magnets 
%72.25.Mk   Interfaces spin polarized transport through

\date{\today}

\maketitle
%Introduction
Molecular based systems are fascinating, yet promising candidates for nanoscale spintronic devices and open viable routes toward quantum computing\ \cite{nmat2133,*410789a0}. Previous experiments on the spin transport through break junctions\ \cite{nature00791} and spin valves\ \cite{Drew2009} unveil exciting new frontiers of molecular magnetism. Much effort is dedicated to understand the properties of organic/magnetic interfaces. To this end, spatially averaging techniques show substantial spin-injection into, as well as long spin-coherent transport throughout films of metal-phthalocyanines (MPc)\ \cite{Suzuki2002,*jp0204760,nmat2334}. However, detailed and quantitative access to different constituents of a single molecule is desirable, though challenging. Scanning tunneling microscopy (STM) is well established as a probe of a local spin\ \cite{Meier2008,science1199,PhysRevLett_99_067202,PhysRevLett_101_197208,*PhysRevLett.103.257202,PhysRevLett_102_167203,PhysRevLett_103_087205,AidiZhao09022005,PhysRevLett.102.257203} in an atomically well defined environment.

Iacovita \textsl{et al.}\ recently performed a spin-polarized STM (SP-STM) study of a CoPc in contact with a ferromagnetic cobalt nano-island\ \cite{PhysRevLett_101_116602}. Stacking contrast, spin-dependent scattering, edge states, mesoscopic relaxations as well as the adsorbate induced modification create a complex environment\ \cite{PhysRevLett.92.057202,*PhysRevLett_96_237203,*PhysRevLett.99.246102,*PRB_77_035417_2008,*scienc_843} toward understanding the influence of the substrate on molecular magnetism. After careful selection of electronically equivalent Co nano-islands a ferromagnetic exchange interaction between the molecular spin and the cobalt lead was successfully deduced, both theoretically and experimentally.

In this letter we demonstrate a significant spin-polarization for a CoPc molecule in contact with a ferromagnetic Fe thin film due to molecule-substrate hybridization even though the molecule loses its net spin. As confirmed by SP-STM, an energy/site-dependent spin polarization from inversion to amplification is resolved on the sub-molecular scale. State-of-the-art density functional theory (DFT), which includes the decisive role of van-der-Waals (vdW) interactions, reveals both the magnetic and electronic nature of the molecule coupled to the ferromagnetic substrate. Even though the net spin of the molecule is lost due to a transfer of one electron, spin-splitting is recovered through the local bonding of molecular orbitals with Fe $3d$ bands.

%Setup

Simulations were carried out in the DFT\ \cite{PhysRev.136.B864} formalism with a plane wave implementation as provided by the VASP code\ \cite{Kresse1994,*Kresse1996}. Pseudopotentials used were generated with the projector augmented wave method\ \cite{Blochl1994} by using the PBE generalized-gradient exchange-correlation energy functional\ \cite{Perdew1996} (GGA). A slab consisting of two Fe and three W atomic layers, with a ($5\times7$) in-plane surface unit cell modeled the molecule-surface system. The kinetic energy cutoff of the plane waves was set to 500~eV while the Brillouin zone was sampled by the $\Gamma$ point. Optimized molecule-surface geometries were obtained by relaxing all molecular degrees of freedom and those of the Fe overlayers by including long-range vdW interactions in a semi-empirical way\ \cite{Grimme2006,Atodiresei2009}. The threshold of the calculated forces was fixed to $0.001~\text{eV}/\text{\AA}$. 

Experiments were performed in an ultra-high vacuum system with an STM operated at 6 K\ \cite{RevSciInstrum_68_3806}. Chromium coated tungsten tips with out-of-plane spin sensitivity\ \cite{PhysRevLett.88.057201} were utilized for all measurements. $1.8~\text{ML}$ Fe was deposited on a W(110) single crystal kept at 400K. Afterward, the magnetic sample was cooled down to measurement temperature and spin sensitivity was confirmed by imaging the alternating spin-up ($\odot$) and spin-down ($\otimes$) domains of the $2^{nd}$ layer portion of the Fe film\ \cite{PhysRevLett.88.057201}. CoPc molecules were thermally sublimated from a home-built crucible onto the pre-cooled substrate to ensure spatially well separated molecules.

All STM images were recorded in constant-current mode at the set current $I$ and sample bias voltage $U$. The local spin-polarization of the atomic-scale tunnel junction can directly be determined from height variations $\Delta s$ resulting from different magnetization directions of electronically identical areas as introduced by Wiesendanger \textsl{et al.}\ \cite{PhysRevLett.65.247}:
\begin{equation*}
%\label{eq8}
P_{\text{t}} P_{\text{s}} \cos{\theta}=\frac{e^{\nicefrac{a \sqrt{\phi} \Delta s}{2}}-1}{e^{\nicefrac{a \sqrt{\phi} \Delta s}{2}}+1}=\tanh{a \sqrt{\phi} \Delta s}\overbrace{\approx a \sqrt{\phi} \Delta s}^{\Delta s < 1~\text{\AA}}.
\end{equation*}
The polarization of tip ($P_{\text{t}}$) and sample ($P_{\text{s}}$) are defined as the normalized difference of the energy integrated spin-up and spin-down states (${P_{{\text{t}}({\text{s}})}} := {\rho}^\odot_{{\text{t}}({\text{s}})} - {\rho}^\otimes_{{\text{t}}({\text{s}})}; {\rho}^\odot_{{\text{t}}({\text{s}})} + {\rho}^\otimes_{{\text{t}}(\text{s})}=1$). ${\theta}$ is the angle between the respective magnetization directions, $\phi$ the mean local tunneling barrier height, and $a\cong 1~\text{eV}^{-\frac{1}{2}}\text{\AA}^{-1}$.

\begin{figure}
\includegraphics[width=0.45\textwidth,clip]{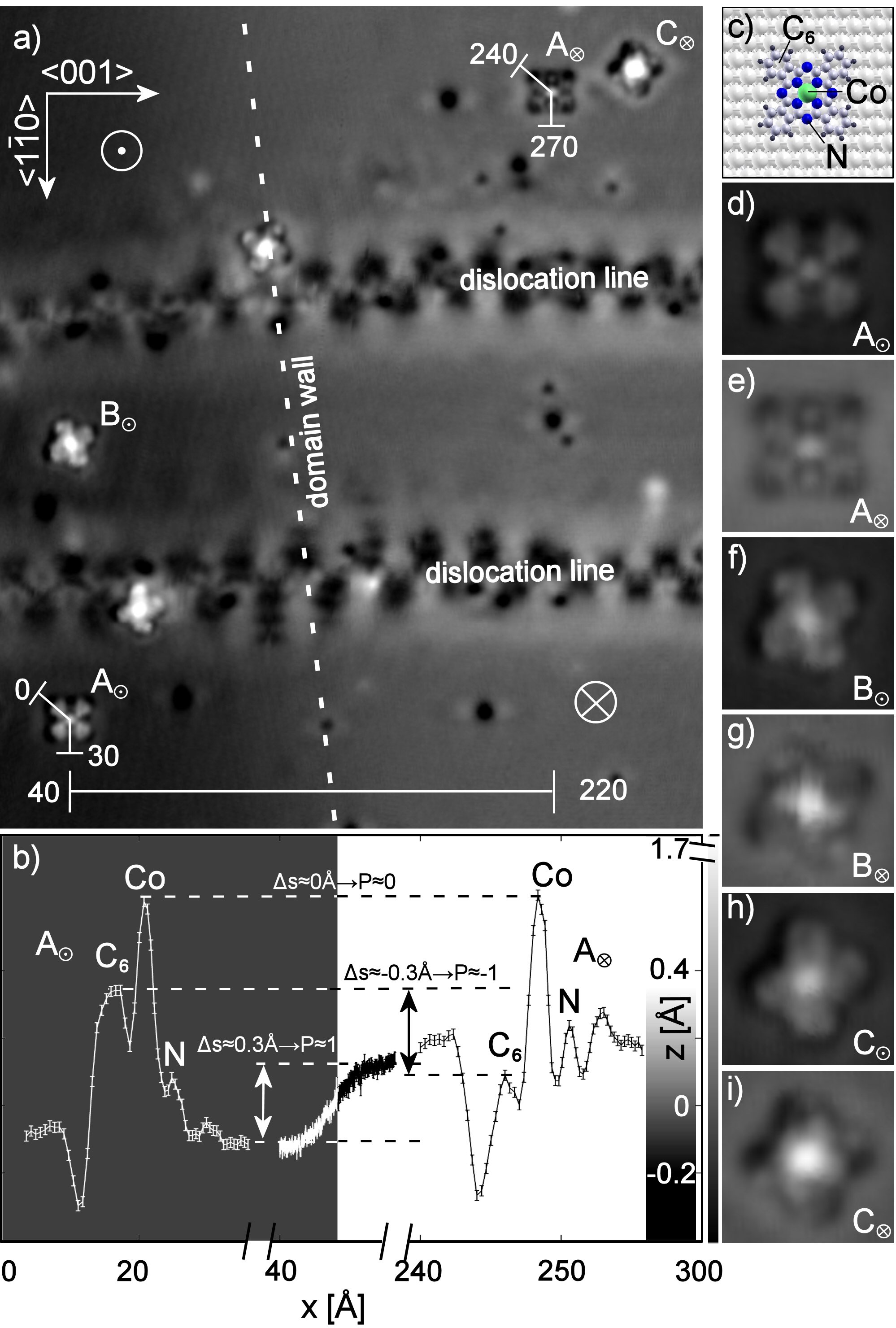}
\caption{\label{fig1}(a) SP-STM image of CoPcs on magnetic iron domains ($\odot,\otimes$). (b) Line profile as indicated in (a). (d)-(i) High-resolution images of CoPcs in all three orientations (A, B, C) on oppositely magnetized domains. (c) Adsorption configuration of a CoPc in orientation A. (a) and (d)-(i) 0.05~V, 200~pA; a) $240~\text{\AA}\times 280~\text{\AA}$; (d)-(i) $25~\text{\AA}\times 25~\text{\AA}$. The colorscale for (a) is non-linear to illustrate the domain structure, while the colorscale for (d)-(i) is linear.}
\end{figure}

%Discussion Overview Image

Figure\ \ref{fig1}(a) shows a representative $2^{nd}$ ML area of Fe after deposition of CoPc molecules. The two dislocation lines indicate the crystallographic axes. Isolated CoPcs adsorb in three well-defined orientations (denoted as A, B, and C). The line profile (\fig{fig1}(b)) along the indicated positions in (a) quantitatively illustrates the height variation within the atomically flat Fe film ($\Delta s=0.3~\text{\AA}$) and the domain-dependent molecular appearance across a type A CoPc on an up- and down-domain. This variation is due to the relative alignment of tip and sample magnetization being either parallel or anti-parallel. While the center of the molecule (Co-site) has the same apparent height on both domains ($\Delta s=0~\text{\AA}$), we see a profound contrast at the ligand-site ($\Delta s=-0.3~\text{\AA}$). For an up-domain the ligand exhibits an increased apparent height relative to the surrounding Fe film, which is in contrast to the apparent height for a molecule on the down-domain.

Thereby, orientation A\ \footnote{Orientation A is only observed when prepared on a pre-cooled surface, indicating a metastable configuration.} represents a highly symmetric configuration (C$_{2v}$-symmetry) with one molecular axis oriented exactly along the $<$001$>$ direction. While there are differences in intramolecular contrast amongst the three molecular orientations (\fig{fig1}(d) to (i)), all three configurations show similar domain dependent contrast variations. In the following, we focus on molecules in configuration A which are not in the vicinity of dislocation lines or domain walls. As the molecular appearance controllably switches under an external magnetic field applied\ \cite{EPAPS}, the magnetic contrast can unambiguously be attributed to the spin-polarized local density of states (SP-LDOS) of the molecule-surface system.

\begin{figure}
\includegraphics[width=.45\textwidth,clip]{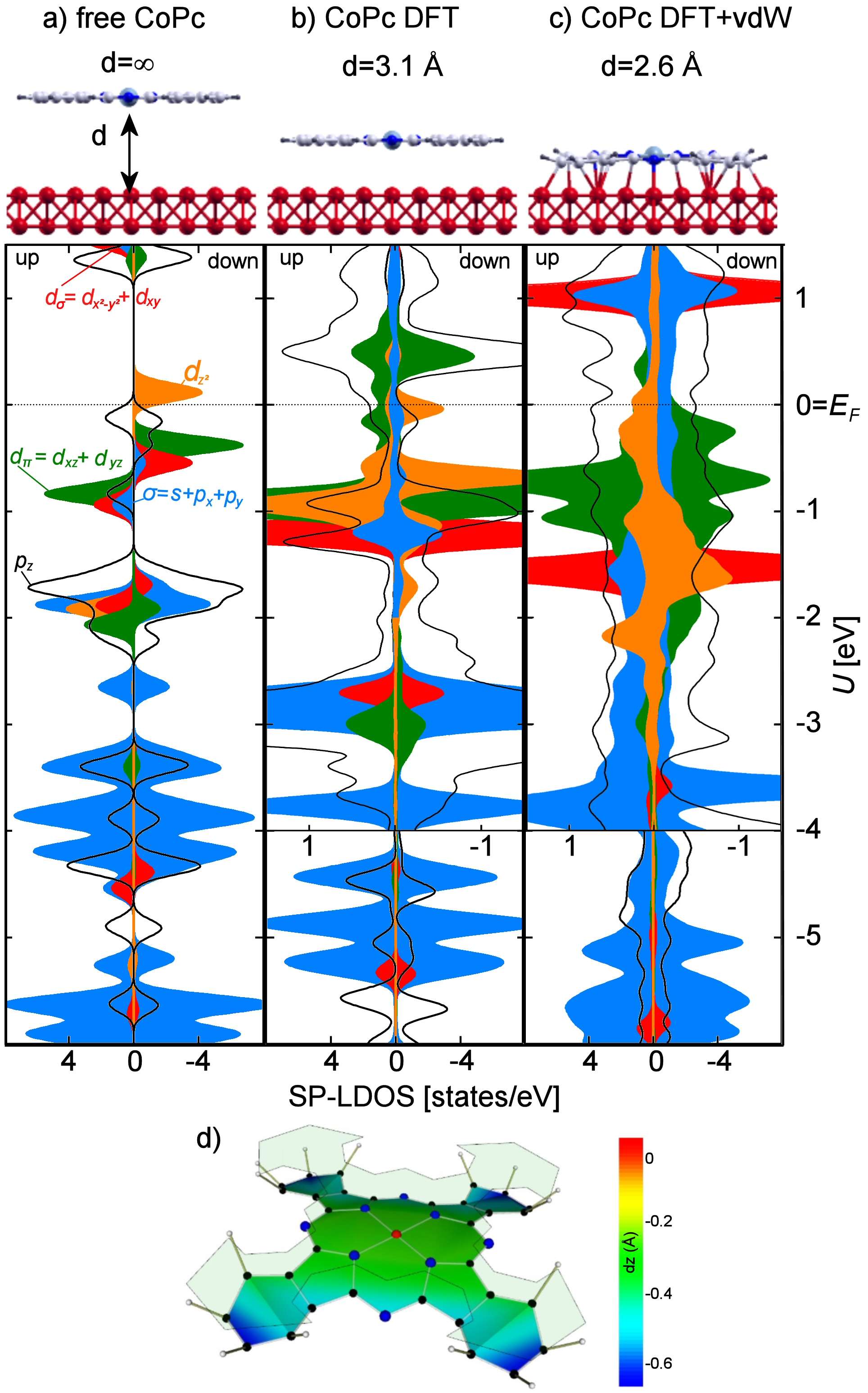}
\caption{\label{theory}The geometry and electronic structure for a free CoPc (a) and adsorbed on an Fe surface (b) without and (c) with  vdW forces included during the relaxation. (d) Molecular deformation due to vdW forces. The C$_6$ rings and two of the outer N are twisted by up to 0.3~{\AA} toward the surface and away from the plane defined by the Co ion and four inner N. All H are pointing away from the surface.}
\end{figure}

%Discussion Theory

First-principles calculations within DFT (GGA) clarify the origin of the observed spin-dependence. Calculations for the A geometry show the preferred top adsorption, i.e.\ the Co atom on top of a surface Fe atom. Previous results for a Co island revealed an adsorption at the bridge site\ \cite{PhysRevLett_101_116602}, while the hollow site was deduced for a Au(111) substrate\ \cite{AidiZhao09022005}.
Furthermore, calculations cover both spin-orbit coupling (SOC) as well as the considerable role of vdW interactions. Results without SOC are discussed in terms of molecular orbitals (MOs) to clearly illustrate the effect of vdW interactions alone.

Figure\ \ref{theory} depicts the change in electronic structure of a CoPc with increasing molecule-substrate interaction. Figure \ref{theory}(a) shows the SP-LDOS for the free molecule. In agreement with previous work\ \cite{ja900484c,ic00010a015,AidiZhao09022005,PhysRevLett_101_197208,*PhysRevLett.103.257202} the origin of spin-splitting is an unpaired electron in a MO with $d_{z^2}$ contribution situated at the Co-site and the total molecular spin $S$ is $\nicefrac{1}{2}$.
Figure\ \ref{theory}(b) presents the effect of the surface, when vdW forces are neglected during the relaxation process. The molecule adsorbs 3.1~{\AA} above the surface remaining flat. The SP-LDOS shows a hybridization of MOs and substrate $3d$ states. According to the spatial extension of the MOs perpendicular to the molecular plane (i.e., with $\pi$ character), the hybridization of MOs containing the Co $d_{z^2}$ atomic orbital is the strongest, followed by those including $d_\pi$ and $p_z$ atomic contributions. The $d_\sigma$ and $\sigma$ MOs are only slightly broadened, compared to the free molecule case, as they are localized in the molecular plane. The spin-splitting of the molecule-surface hybrid states is reduced due to a transfer of an electron from the substrate to the CoPc. As a result the formerly unoccupied $d_{z^2}$ type MO becomes occupied and the total molecular spin is quenched (0 $\mu_b$).

The role of vdW forces is crucial as it brings the molecule 0.5~{\AA} closer to the surface (\fig{theory}(b)) and distorts the molecular geometry (\fig{theory}(d)). This new adsorption geometry has a drastically different electronic structure due to the overall changes of hybridization between molecule and substrate.
The (SP-LDOS \fig{theory}(c)) shows not only MOs with a $\pi$ character strongly hybridize with spin-polarized Fe $3d$  states of the same symmetry to form broad spin-split bands but also $\sigma$ type MOs are significantly affected by the interaction with the surface. Again, a transfer of one electron from the surface to the molecule annihilates the molecular spin, but spin-splitting is recovered due to the local molecule-surface bonding at different parts of the molecule. The newly formed molecule-surface hybrid states have, within a given energy interval [$E,E'$], an unbalanced, locally varying electronic charge in the up and down channels which is mapped in SP-STM ($E=E_F, E'=E_F+eU$). %The spin splitting is only recovered when considering the correct vdW relaxed molecule-surface geometry and properly accounts for the observed spin-contrast in experiment.

\begin{figure}
\includegraphics[width=.45\textwidth,clip]{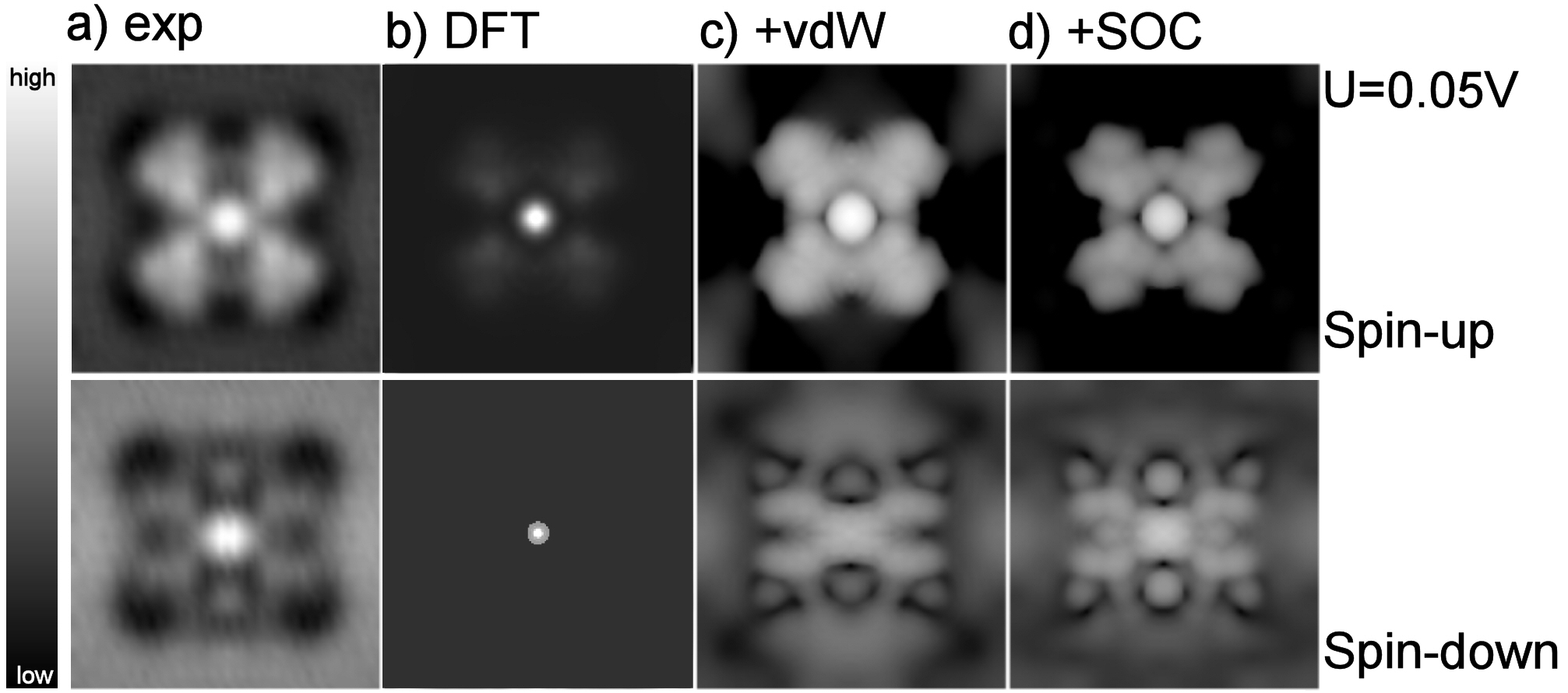}
\caption{\label{comparison}Experimental and simulated SP-STM images at $U=0.05~\text{V}$ and $22~\text{\AA}\times 22~\text{\AA}$ for both spin directions. (a) Averaged experimental images. (b)-(d) Isocharge surfaces;\ (b) conventional DFT (GGA) (without vdW), (c) DFT+vdW, (d) DFT+vdW and including SOC.}
\end{figure}

For a direct comparison of the first principles calculations with constant-current SP-STM images, isocharge surfaces above the CoPc are extracted from the spatial variation of the energy integrated SP-LDOS. This approach mimics the experimental situation of a local and spin-sensitive tip probing the charge density above the surface at constant current and thereby, accounts for the variation of decay lengths of the different states into vacuum. Figure\ \ref{comparison} gathers calculated spin dependent isocharge surfaces and experimental SP-STM images for $U=0.05~\text{V}$. The isocharge surfaces reproduce all features of the experimental SP-STM images, only if vdW forces forces are taken into account. Corrections due to the inclusion of spin-orbit coupling effects are minor.

%
%Discussion Theory vs Exp
%
\begin{figure}
\includegraphics[width=0.45\textwidth,clip]{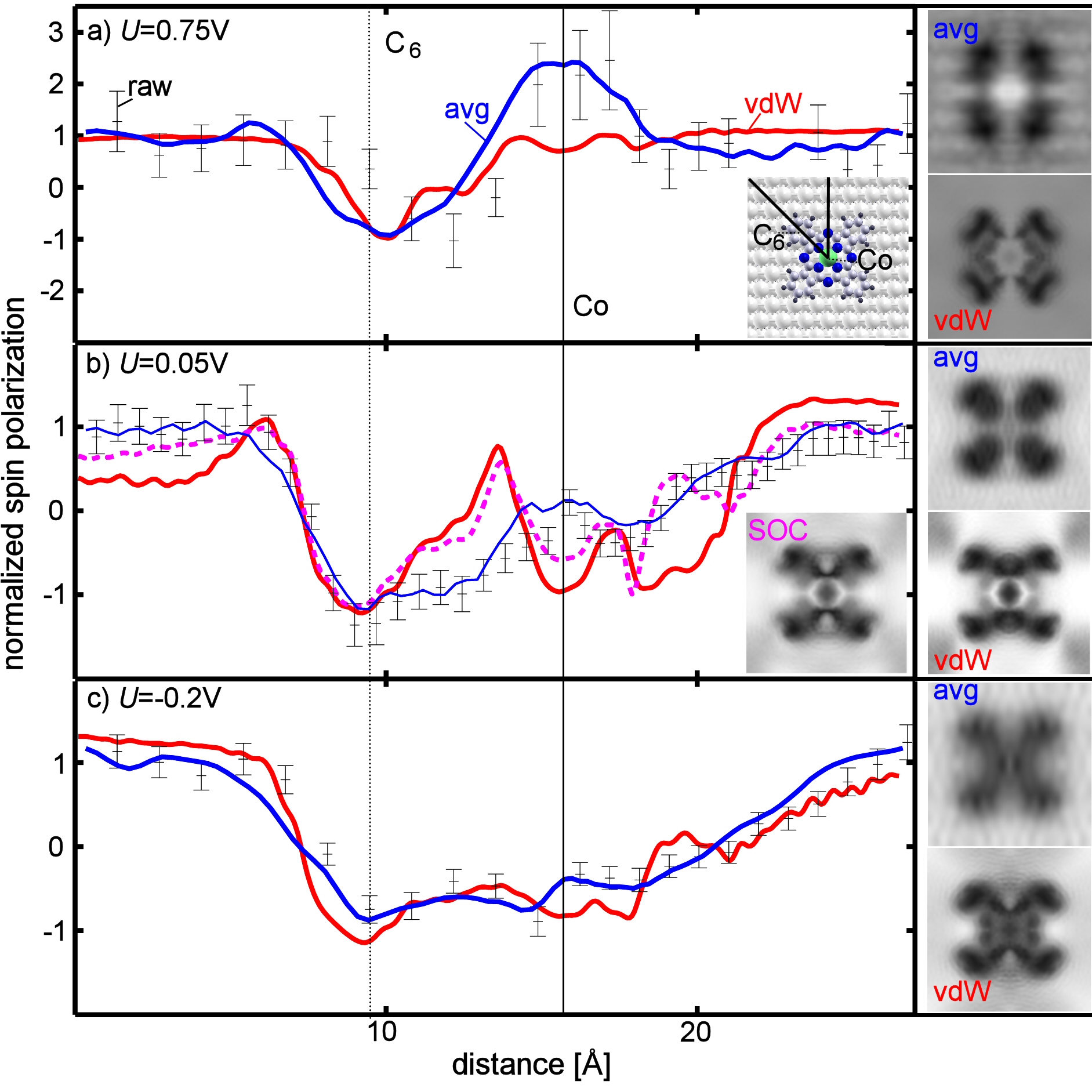}
\caption{\label{fig4} Local spin polarization at (a)-(c) three representative energies. Raw- and  {\color{blue}avg} (over multiple images and according to symmetry\ \cite{EPAPS}) experimental data are compared with DFT simulations incl.\ {\color{red}vdW} and incl.\ {\color[rgb]{1,0,1}{S}}pin-{\color[rgb]{1,0,1}{O}}rbit-{\color[rgb]{1,0,1}{C}}oupling. Line profiles follow high-symmetry directions as indicated in the sphere model inset in (a). Simulated data including SOC is only given in (b) as the SOC corrections for (a) and (c) are negligible. Insets: $22~\text{\AA}\times 22~\text{\AA}$.}
\end{figure}

Figure\ \ref{fig4} illustrates quantitatively the local spin polarization. To compensate for the tip polarization, we normalized the experimental polarization above the molecule ($P^\text{exp}_\text{CoPc}$) relative to the spin polarization of the free surface $(P^\text{exp}_\text{Fe})$,
\begin{equation}
\label{eq9}
P=\frac{P^\text{exp}_{CoPc}}{P^\text{exp}_{Fe}}=\frac{P_{\text{t}} {P}_\text{CoPc} \cos{\theta}}{P_{\text{t}} {P}_\text{Fe} \cos{\theta}}=\frac{{P}^{sim}_\text{CoPc}}{{P}^{sim}_\text{Fe}}.
\end{equation}
The normalized polarization ($P$) is therefore, to a first approximation, independent of the tip polarization and can directly be compared to our simulated polarization (${P}^{sim}$). 

For comparison both experimental and theoretical spatial maps of the polarization at three characteristic energies are given in \fig{fig4}. We see an excellent agreement of experimental findings and DFT calculations, when the vdW relaxed adsorption geometry is used. Remarkably, not only reduction ($1 > P > 0$) but also inversion ($P < 0$) and amplification ($P > 1$) of spin polarization can equally be observed. The $P$ above the C$_6$-ring shows an inversion for all energies, while the spin polarization near the Co-site ranges from inversion (at -0.2~V) to amplification (at 0.75~V). %A similar sign reversal is predicted for CoPc/Co/Cu\ \cite{PhysRevLett_101_116602}, but was experimentally not resolved. There, the discrepancy has been tentatively discussed in terms of SOC and transport effects. We note, that in the analysis of magnetic asymmetry in terms of spin polarization it is a prerequisite that the tip-sample distance for oppositely magnetized molecules at given stabilization parameters is the same, or the difference must be accounted for by additional model assumptions\cite{Kubetzka2003,Yamada2005}. The experimental approach used in this work does not suffer from this drawback.

%
%Conclusions
%
In conclusion we investigated single CoPc molecules adsorbed on a ferromagnetic Fe surface and locally observed the energy-dependent spin transport with intramolecular spatial resolution. The tunneling current above the entire molecule - metal ion, as well as the organic ligand - shows a high, locally varying spin polarization ranging from inversion up to amplification with respect to the ferromagnetic Fe film. Optimized molecule-surface geometries for the first principles calculations were obtained by including long-range van-der-Waals interactions during the relaxation, leading to an excellent agreement of theoretical and experimental data. Therefore, the observed spin polarization is identified as a unique property the molecule-substrate hybrid states, by our combined DFT and SP-STM approach.

\begin{acknowledgments}
This work is funded by the DFG (SFB 668-A5, SPP1243 and GrK 611), the JSPS, Alexander von Humboldt foundation, the EU project SpiDME, the ERC Advanced Grant FURORE, and the City of Hamburg (Cluster of Excellence NANOSPINTRONICS).
\end{acknowledgments}

%

%\bibliographystyle{ghapsrev4-1}
%\bibliography{library}

\end{document}